\PassOptionsToPackage{table,xcdraw}{xcolor}
\documentclass{aastex6}
\usepackage{natbib}
\usepackage{graphicx}
\usepackage{color}
\usepackage{amsmath}
\usepackage{epstopdf}
\begin{document}
	
\title{An Analysis of Stochastic Jovian Oscillation Excitation by Moist Convection}
	\author{Ethan Dederick}
	\affil{New Mexico State University}
	\email{dederiej@nmsu.edu}
	
	\author{Jason Jackiewicz}
	\affil{New Mexico State University}
	\email{jasonj@nmsu.edu}
	
	\author{Tristan Guillot}
	\affil{Universit\'{e} de Nice-Sophia Antipolis, Observatoire de la C\^{o}te d'Azur}
	\email{tristan.guillot@oca.eu}
	
	\maketitle
	
	\section{Abstract}

        Recent observations of Jupiter have suggested the existence of global oscillatory modes at millihertz frequencies, yet the source mechanism responsible for driving these modes is still unknown. However, the energies necessary to produce observable surface oscillations have been predicted. Here we investigate if moist convection in Jupiter's upper atmosphere can be responsible for driving the global oscillations and what moist convective energy requirements are necessary to achieve these theoretical mode energies and surface amplitudes. We begin by creating a one-dimensional moist convective cloud model and find that the available kinetic energy of the rising cloud column falls below theoretical estimates of oscillations energies. That is, mode excitation cannot occur with a single storm eruption. We then explore stochastic excitation scenarios of the oscillations by moist convective storms. We find that mode energies and amplitudes can reach theoretical estimates if the storm energy available to the modes is more than just kinetic. In order for the modes to be excited, we find that they require $5 \times 10^{27}$ to 10$^{28}$ erg per day. However, even for a large storm eruption each day, the available kinetic energy from the storms falls two orders of magnitude short of the required driving energy. Although our models may oversimplify the true complexity of the coupling between Jovian storms and global oscillations, our findings reveal that enough thermal energy is associated with moist convection to drive the modes, should it be available to them.

	\section{Introduction}
	Since the advent of helioseismology and the detection of solar global oscillations, it has been postulated that gas giant planets could also exhibit similar oscillations. Much theoretical work has been done investigating the inherent nature of the oscillatory modes of Jupiter and Saturn \citep[e.g.,][]{Vorontsov1976,Bercovici_Schubert1987,Marley_Porco1993,Jackiewicz2012}, and indeed, they have been observed. \citet{Gaulme2011} and \citet{Hedman_Nicholson2013} have detected signatures of Jupiter's global oscillations and the spiral density structures in Saturn's rings caused by Saturnian fundamental modes, respectively. Yet the excitation mechanism of these oscillations remains a mystery. In the case of the Sun, \citet{Goldreich_Keeley_2_1977} demonstrated that the energy from turbulent convection drives the observed solar oscillations. However, a similar approach on Jupiter reveals that too little energy is available from convection to be responsible for the global oscillations \citep{Gaulme2014}. The ratio of velocity of the convective flux to the sound speed (Mach number) gives an indication as to the energy in a driving mechanism. In Jupiter, the Mach number relative to the Sun is much lower, resulting in oscillatory surface velocity amplitudes of at least three orders of magnitude less than the Sun ($\le 0.5$ m s$^{-1}$ on Jupiter compared to $\sim 500$ m s$^{-1}$ on the sun) \citep{Bercovici_Schubert1987}. While \citet{Bercovici_Schubert1987} claim that turbulent motions may be responsible for oscillatory surface velocities of $\sim50$ cm s$^{-1}$, these turbulent motions may not be the result of traditional convection.

        Several possible sources have been investigated in regard to their ability to drive Jovian global oscillations but all were found to be insufficient. Both the $\kappa$-mechanism, an opacity effect traditionally observed in pulsating variable stars \citep{Cowling1941,Stothers1976}, and the radiative suppression mechanism,  a possible excitation source for hot Jupiters, were shown to be unable to drive Jovian oscillations by \citet{Dederick_Jackiewicz2017}. Although Helium rain, the process by which atomic Helium droplets rain through the immiscible metallic hydrogen region, has yet to be examined as an excitation mechanism, we think this process is unlikely to excite modes as the inertia at these pressures is too large for driving to occur. Ortho- to para-hydrogen conversion as a driving mechanism is unlikely as well as it is a very slow process, too slow to couple with oscillatory motions \citep{Bercovici_Schubert1987}. The mechanism that excites Jovian oscillations has yet to be determined. In this investigation, we examine moist convection and the energy it could supply to drive Jupiter's global oscillations.

	 In the simplest of terms, moist convection is the process of the formation and dissipation of thunderstorms. The main difference between moist convection on Earth and  on Jupiter is the ratio of the molecular weight of water to the surrounding atmospheric gases. On Earth, with the principle atmospheric constituent being diatomic nitrogen, this ratio is less than one, but on Jupiter, being comprised of mainly hydrogen and Helium, the ratio is greater than one. Thus, on Jupiter, the water vapor tends to condense and sink whereas it remains aloft on Earth. A more in-depth description of the Jovian moist convective process is given in the context of our model in Section~\ref{CloudModel}.

         In Section~\ref{CloudAnalysis} we discuss if moist convection can excite Jovian oscillations with a single storm using our cloud model. Section~\ref{HarmonicOsc} contains the explanation of our harmonic oscillator model and Section~\ref{HarmonicResults} lists the results of the model. In Section~\ref{Discussions} we discuss whether moist convection can excite the modes stochastically. Finally we end with some conclusions in Section~\ref{Conclusions}.
	 
	 \section{Cloud Model}\label{CloudModel}
	 In the moist convective cloud analysis we adopt the entraining jet model proposed by \citet{Stoker1986} with a few assumptions. First, we estimate that the moist convection zone on Jupiter begins at the cloud base around 6 bars of pressure \citep{Ingersoll2000}. The maximal height of the cloud is defined as where the vertical cloud velocity drops to zero, which is model dependent. This is partially a consequence of the inferred abundance ratio of water in the deep atmosphere and the assumption that the abundance ratios of O/H, C/H, N/H and S/H are three times solar (see Section~\ref{waterabundance} for a discussion of changing the water abundance on Jupiter). Second, we assume Jupiter's atmosphere is 13.49\% Helium and 86.51\% molecular Hydrogen by particle abundance with trace amounts of heavier gases \citep[taken from Galileo measurements from][]{Niemann1996}. Finally, the entraining jet model assumes a convectively inhibitive water loading process in which the following scenario occurs.

	 As Jupiter has no nuclear fusion processes like the Sun, its convection is driven by cooling at the surface rather than heating from below. As the atmosphere cools, dry parcels of air cool faster than moist parcels (those containing water vapor). This is due to the fact that as moist parcels cool, water condenses and sinks out of the parcel while releasing latent heat. This heat keeps the virtual temperature (temperature of the moist parcels) higher than the atmospheric temperature (temperature of the dry parcels). As long as the virtual temperature is higher than the atmospheric temperature, the atmosphere is stable against convection. (Note: once the water vapor condenses out of a moist parcel, the parcel is still considered with the virtual temperature.) Once all the water vapor has condensed out of the moist parcels and the virtual temperature cools to match the atmospheric temperature, the atmosphere is no longer stratified and thus becomes unstable to convection. At this point, all the condensed water currently resides just below the cloud base at 6-8 bars (6 in our model). However, this region is now much warmer than the air above it (which has continued to cool), and the buoyancy force creates a rising cloud column. This rising cloud is modeled as an entrained jet. That is, the cloud is a rising, convective column of saturated air that entrains, or traps, dry air parcels via convective turbulence as it rises. For a schematic of a moist convective cloud model we refer the reader to Figure 13 in \citet{Fletcher2017}.

	 This column has some temperature profile larger than that of the surrounding atmosphere, and as such, will have a positive vertical velocity associated with it. At this point the cloud can be thought of as either a standalone storm, or, more likely, one cloud column amongst several in a larger storm. The cloud will rise to some height (usually less than 1 bar of pressure), where it will then dissipate its energy into the surrounding environment, expedited by the upper atmospheric winds. As it dissipates, it begins to saturate the dry surrounding air, thereby redistributing its water vapor, starting the entire water loading process over again.
	 
         We model this process using three free parameters:  the relative humidity, defined as the ratio of the vapor pressure to the saturation vapor pressure, the entrainment constant $\alpha$, and the characteristic radius of the cloud column, $r$. The entrainment rate, $\phi$, is the change in mass of dry air in the column per change in pressure per unit mass of moist air. Measuring the entrainment rate of a cloud explicitly would require in-situ measurements. Therefore, we simply approximate the entrainment rate to have the form
	 \begin{equation}
	 \phi = \frac{\alpha}{r}.
	 \end{equation}
	 The entrainment constant, $\alpha$, is a proportionality constant determined from laboratory plume experiments \citep{Stommel1947}. For terrestrial clouds, $\alpha$ takes a value of about 0.2 \citep{Simpson1965}. Therefore, allowing for potential differences in Jovian moist convection, we vary $\alpha$ between 0.01 and 0.4. In addition, we also vary the relative humidity from 0\% to 100\% and $r$ from 0.1 km to 1000 km. 
         % When varying each parameter, we keep the default values at R.H. $= 100\%$, $\alpha = 0.2$, and $r = 1$ km. 
         \citet{HuesoEtAl2002} find that the radius of the cloud column cannot be substantially larger than 25 km as it would violate mass continuity, so radii larger than this are unrealistic. Nevertheless, we run models for $r$ up to 1000~km to fully explore the range of storm sizes on Jupiter. 

One could also assume finite values of the liquid water mixing ratio, $\eta_L$. However, we find its only effect is to decrease maximum cloud height. As more water in the cloud condenses (higher $\eta_L$), the cloud becomes heavier and rains out instead of ascending further. A lower ceiling for the cloud to rise to results in less kinetic energy in the cloud. In the following, kinetic energy refers to the kinetic energy of the rising cloud only. Since  we are interested in a maximum energy scenario, $\eta_L$ is set equal to zero. Finally, for our background atmospheric model the environmental temperature, density, and pressure profiles are taken from \citet{Gaulme_Mosser2005}, which were derived from Galileo data. The derivation of equations used to develop this model can be found in \citet{Stoker1986}. A comprehensive list of the constraints used in this investigation can be found in Table~\ref{Constraints}.

\begin{table}[]
	\centering
	\caption{The list of constraints used in this investigation. The values have been taken or derived from \citet{Gudkova_Zharkov1999,Gierasch2000,Ingersoll2000,HuesoEtAl2002,Gaulme2011}.}
	\label{Constraints}
	\begin{tabular}{|c|c|}
		\hline
		\textbf{Constraint}                   & \textbf{Value}                                         \\ \hline
		Cloud Base                            & $\sim$6 bars                                           \\ \hline
		Maximum Cloud Column Radius           & 25 km                                                  \\ \hline
		Storm Thermal Power                   & $5 \times 10^{22}$ erg s$^{-1}$ \\ \hline
		Storm Area                            & $10^6$ km$^2$              \\ \hline
		Surface Storm Density                 & $0.66 \times 10^{-9}$ km$^{-2}$ \\ \hline
		l=2 n=0 Mode Energy for 1 m Amplitude & $6.4 \times 10^{26}$                        \\ \hline
		l=0 n=1 Mode Energy                   & $10^{26}$ - $10^{27}$ erg      \\ \hline
		Cumulative Mode Surface Amplitude     & 10 - 100 m                                             \\ \hline
	\end{tabular}
\end{table}

	 \section{Analysis}\label{CloudAnalysis}
\begin{figure}[t]
	 	\centering
	 	\centerline{
	 	\includegraphics[width=.5\textwidth]{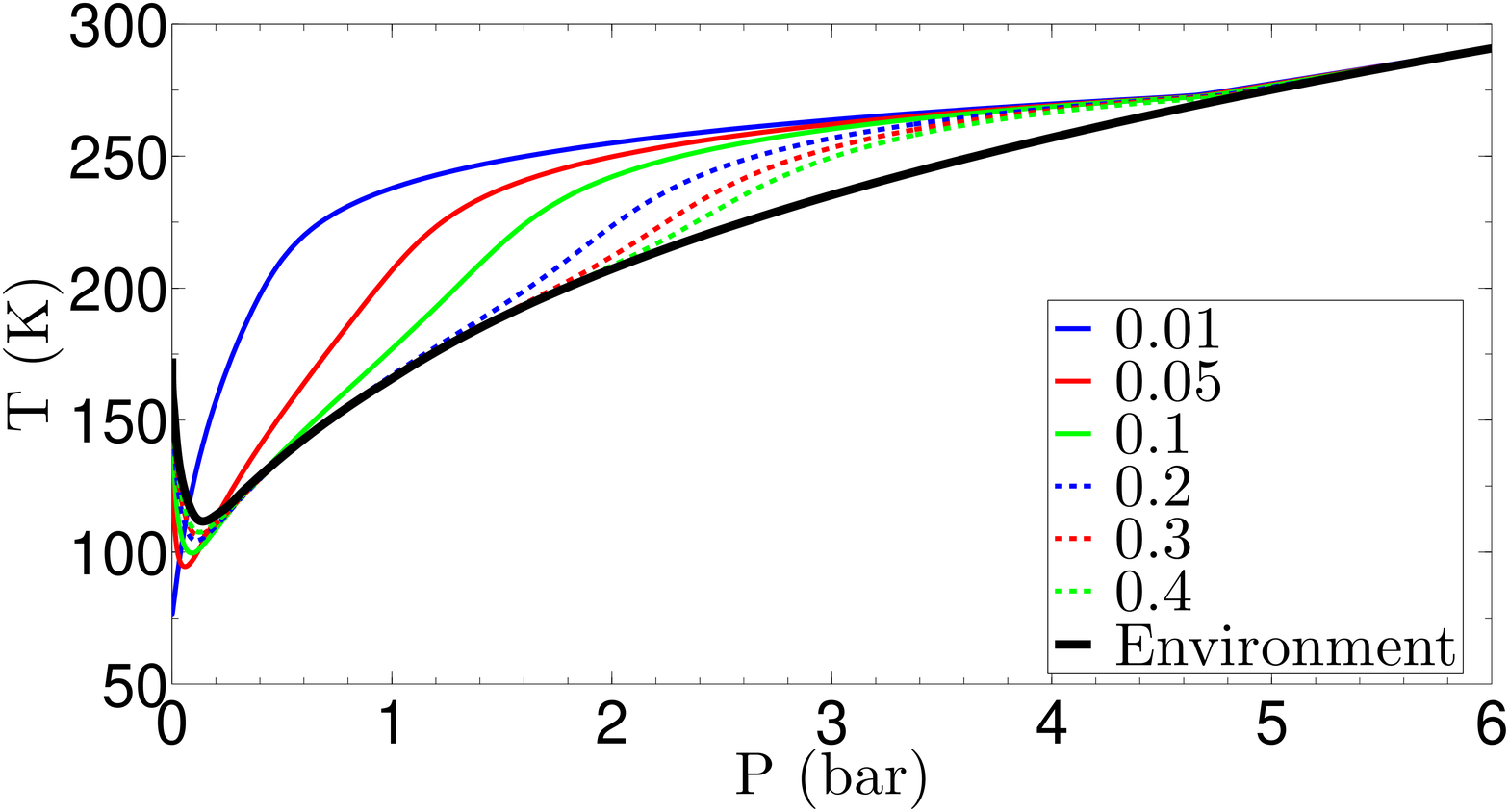}
		\includegraphics[width=.5\textwidth]{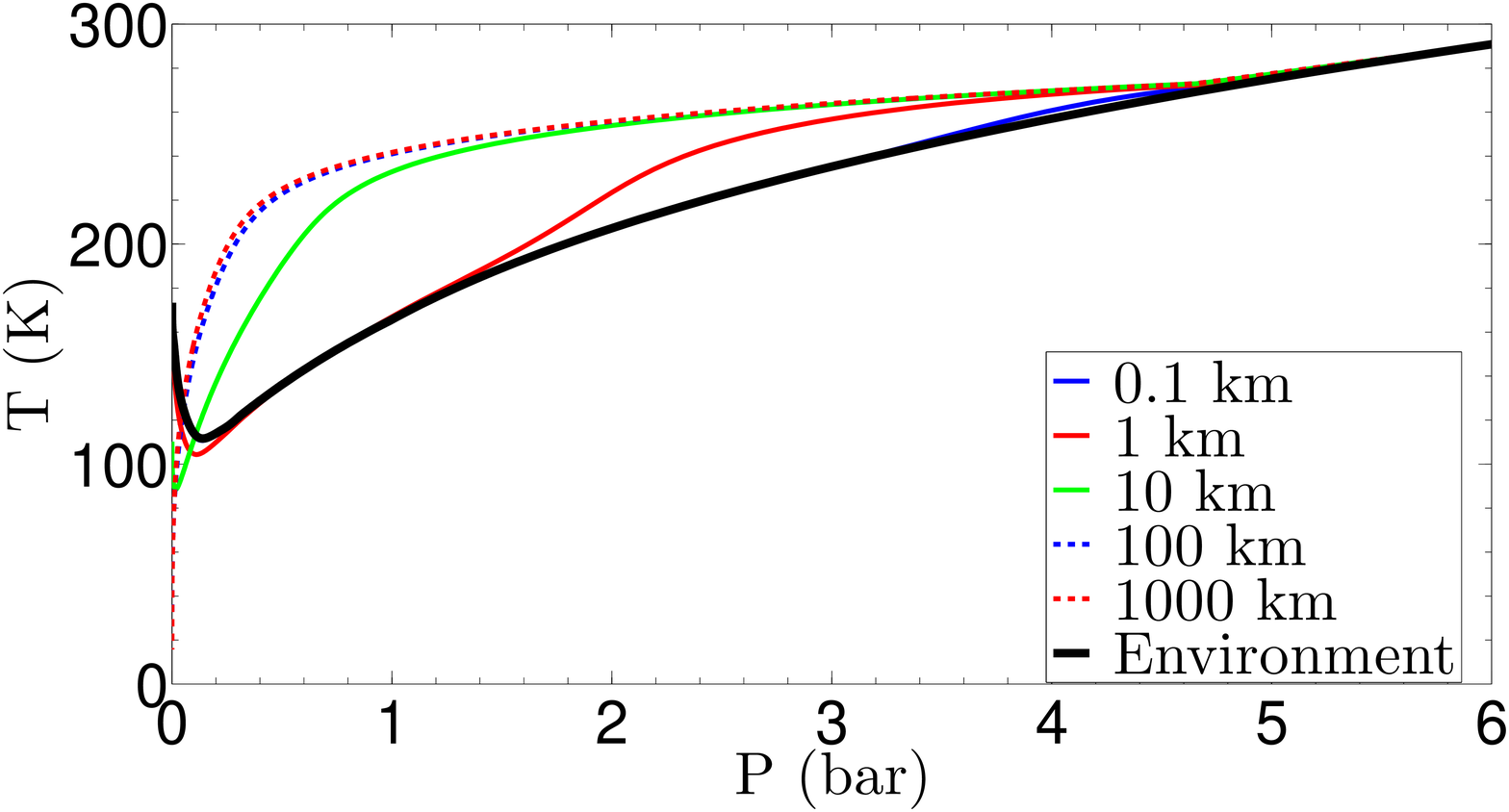}}
	 	\caption{Left: The cloud temperature profile for varying values of the entrainment constant $\alpha$ shown in the legend. The relative humidity is fixed at 100\% and the cloud radius at 1~km. Right: The cloud temperature profile for various radii at $\alpha=0.2$. In both plots the black line represents the temperature profile of the surrounding air from the Jovian atmospheric data.}
	 	\label{fig:CloudT}
	 \end{figure}

	 Figure~\ref{fig:CloudT} shows the temperature profiles of the environment and the cloud column for various entrainment constant values and radii. We do not plot the cloud temperature profiles for varying relative humidity as changes in the relative humidity do not significantly affect the cloud temperature, especially relative to the other two parameters. Note how the clouds are always warmer than the ambient environment. The cloud top can be defined where the cloud temperature equilibrates with the environmental temperature.

	 \begin{figure}[t]
	 	\centering
	 	\centerline{
	 	\includegraphics[width=.5\textwidth]{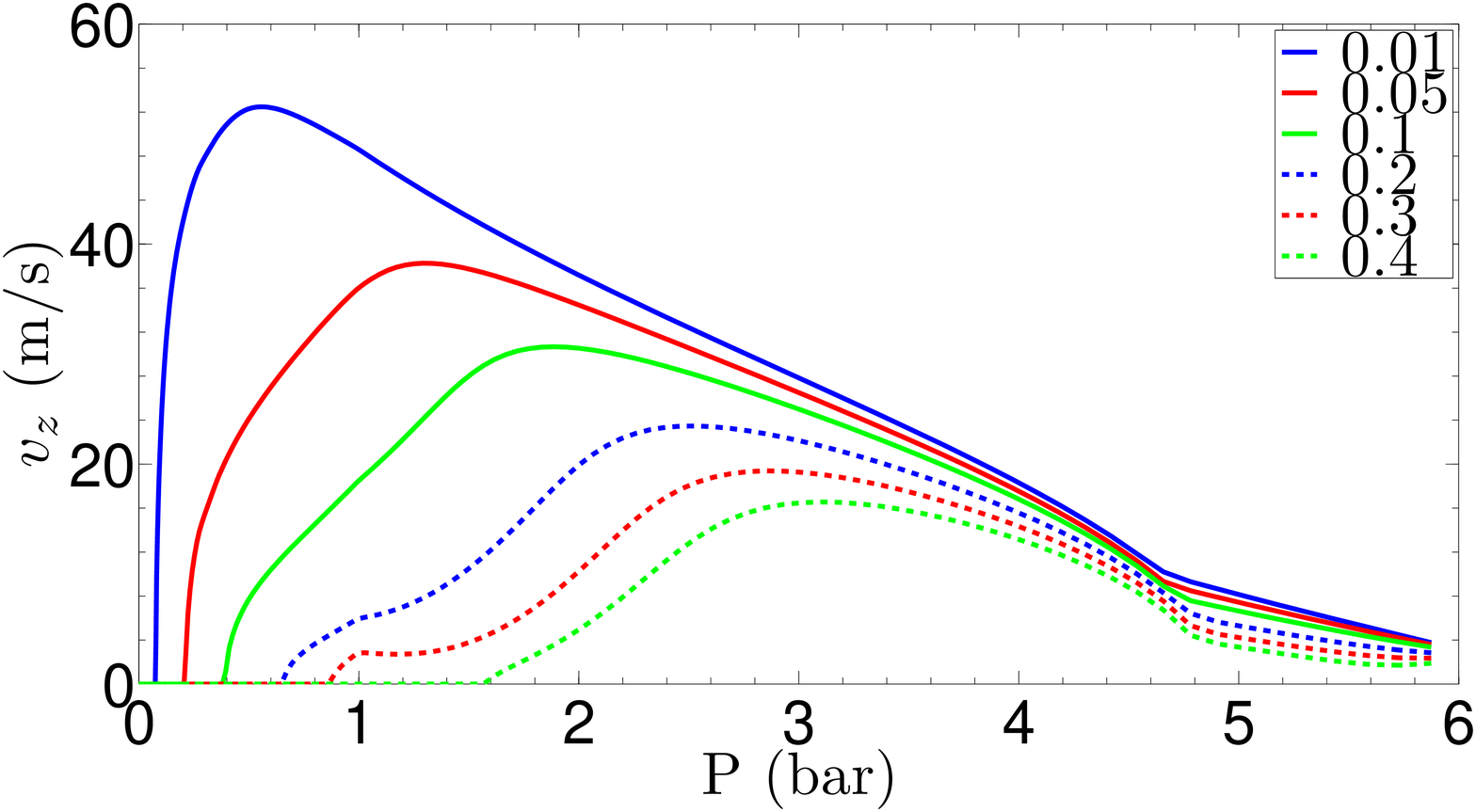}
		\includegraphics[width=.5\textwidth]{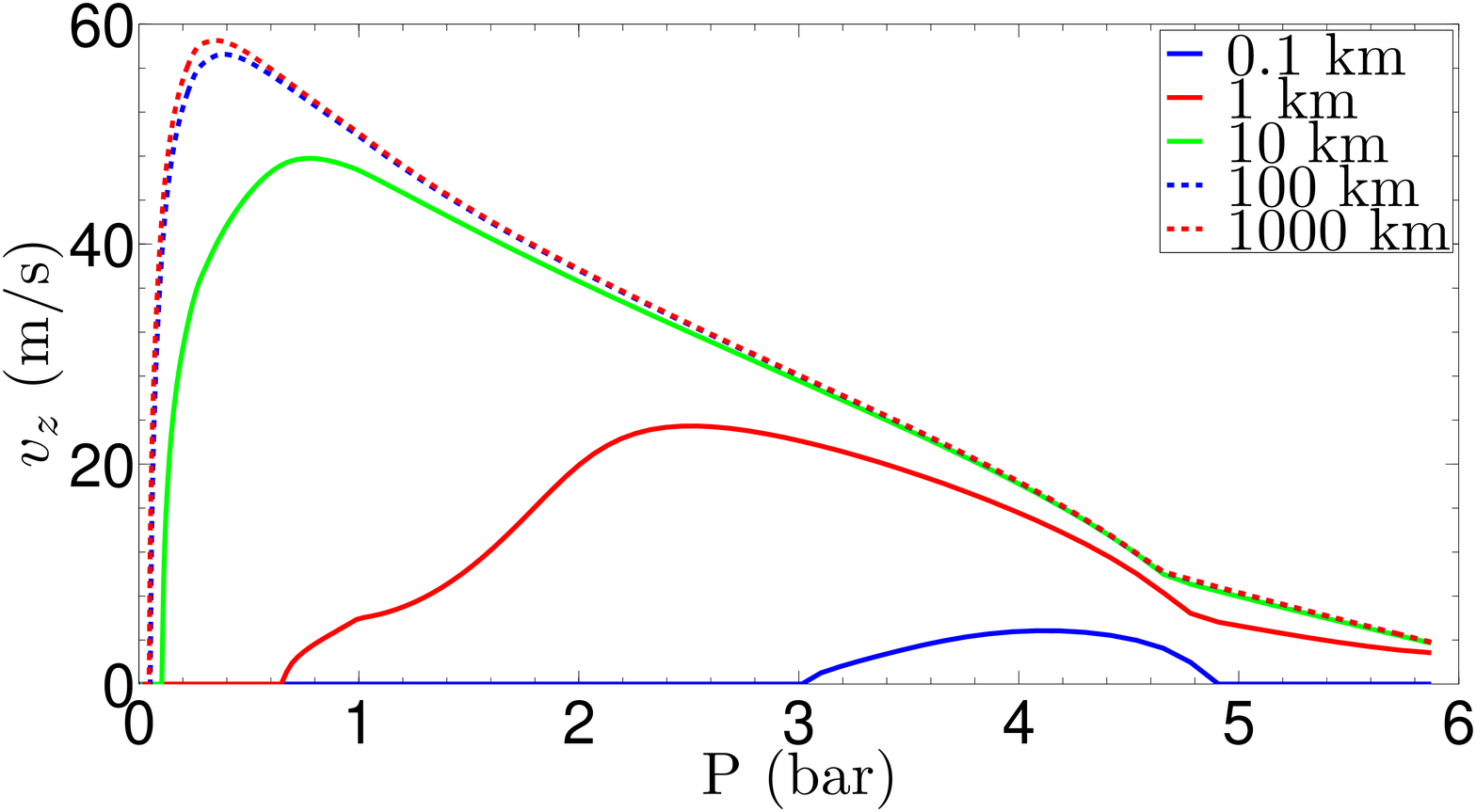}}
	 	\caption{Left: Vertical velocity $v_z$ profiles of the cloud column for varying values of the entrainment constant $\alpha$ shown in the legend. The relative humidity is fixed at 100\% and the cloud radius at 1~km.  Right: Vertical velocity profiles of the cloud column for varying radii at $\alpha=0.2$.}
	 	\label{fig:CloudVel}
	 \end{figure}	 
	 
	 \begin{figure}[t]
	 	\centering
	 	\centering{
	 	\includegraphics[width=.49\textwidth]{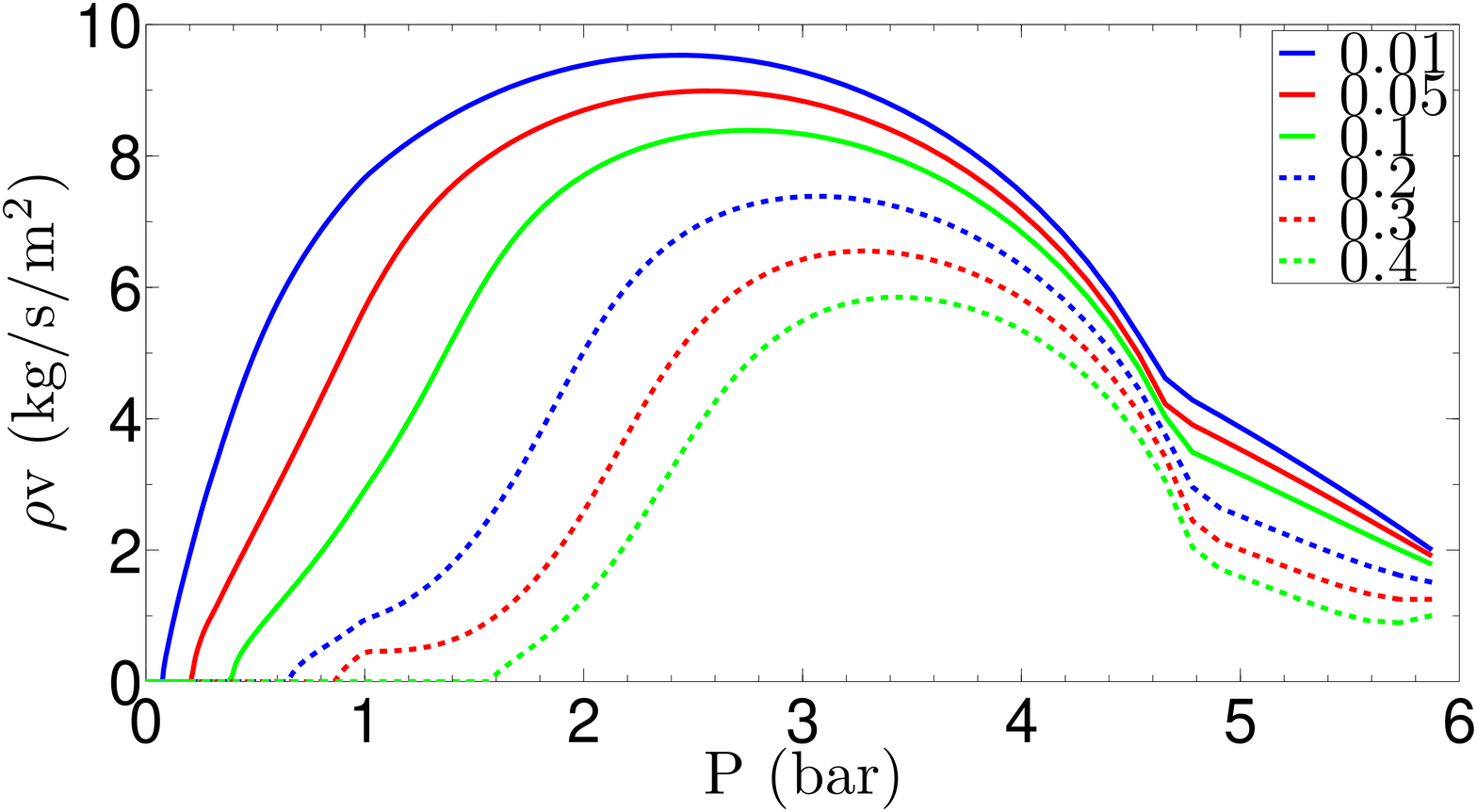}
		\includegraphics[width=.49\textwidth]{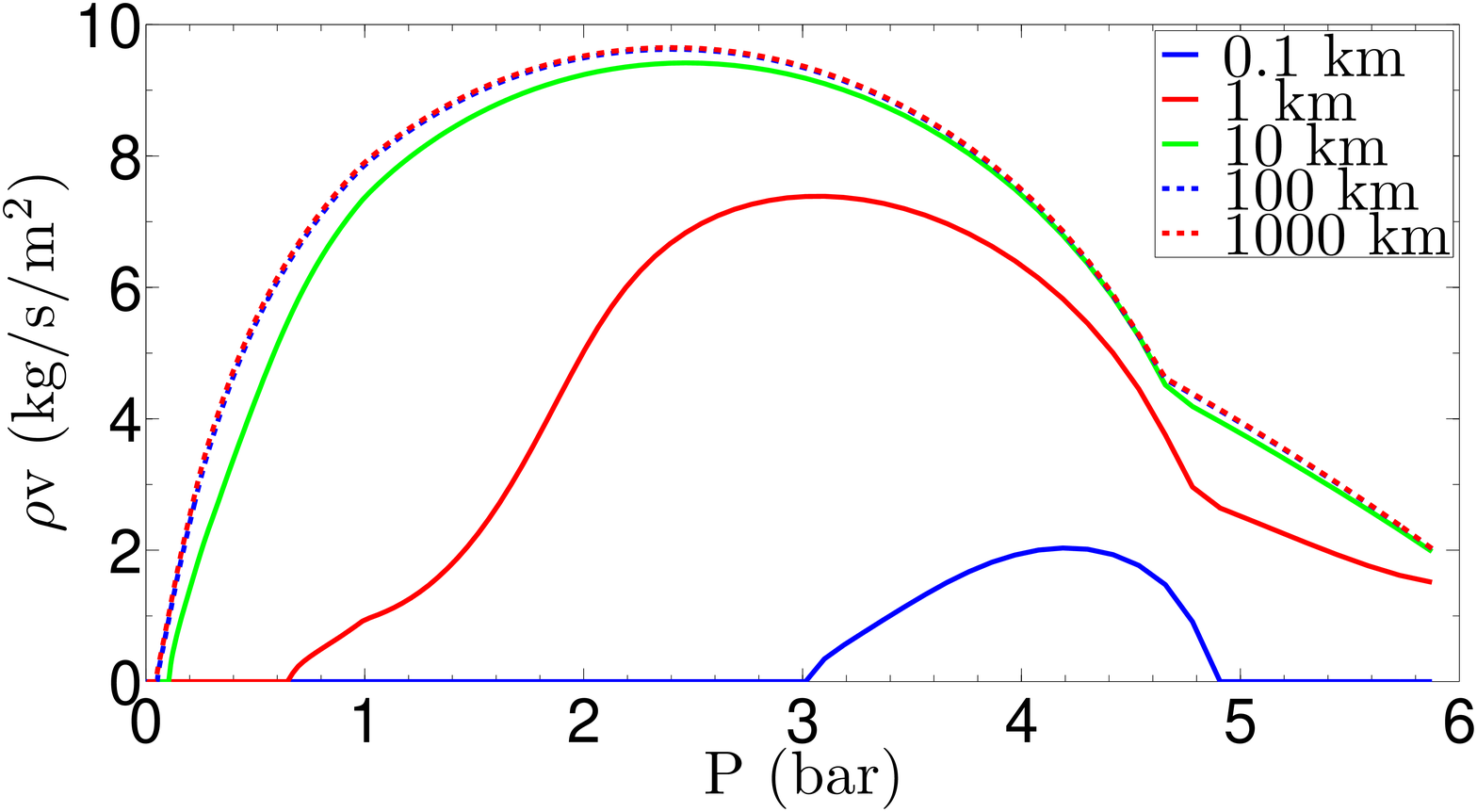}}
	 	\caption{Left: Vertical momentum profiles of the cloud column for varying values of the entrainment constant $\alpha$ shown in the legend. The relative humidity is fixed at 100\% and the cloud radius at 1~km.  Right: Vertical momentum profiles of the cloud column for varying radii at $\alpha=0.2$.}
	 	\label{fig:CloudMomen}
	 \end{figure}

	 A better illustration  of where the cloud top terminates can be seen in Figure~\ref{fig:CloudVel} where  the vertical velocity profiles of the cloud are shown. The velocity is calculated as described in \citet{Stoker1986}. Where the velocity drops to zero also dictates the cloud top layer (ignoring any overshoot).   Finally, Figure~\ref{fig:CloudMomen} shows the vertical momentum profiles ($\rho v_z$), indicating the kinetically energetic regions in the clouds. 

         The moist convection storm column is affected by all of these parameters. $\alpha$, the entrainment coefficient, describes how much the cloud column interacts with the environment. That is, the amount of drag the column feels from rising through the environment. As seen in Figure~\ref{fig:CloudVel}, for a decreasing value of $\alpha$, the cloud extends further towards the Jovian surface; there is less interaction with the environment for lower $\alpha$, and the cloud can extend higher while reaching higher velocities. Low $\alpha$ scenarios are the most probable candidates for overshoot, but are beyond the scope of this paper.

	 The effective cloud column radius, $r$, is also intuitive. For a larger value of $r$, the cloud may have more total interaction with the environment (greater absolute surface area), but it also carries more momentum. As $r$ increases, the momentum to surface drag ratio also increases, resulting in the pattern seen in Figure~\ref{fig:CloudMomen}. Larger column radii cause the cloud to have more momentum and thus rise to higher altitudes. With this interpretation it can be seen that a larger cloud with greater momentum and less drag, or interaction with the environment, will ascend to a lower pressure layer and carry with it more kinetic energy. It is on this basis that we calculate the power produced by these cloud columns.

	 \subsection{Maximum Energy/Power}\label{maxenergypower}
	 The ideal scenario for driving Jovian oscillations occurs when the maximum possible moist convective energy is available and all of it is transferred to the oscillations. In such a scenario, we assume that all the energy of the cloud available for dispersion is kinetic, such that the energy density $U=\rho v^2$. Since we also use a vertical cloud column entrainment model, we can find the kinetic energy per vertical length $dz$ with
	 \begin{equation}
           % E_{dz}=\pi r^2Udz
           dE = \pi r^2 U\, dz.
	 \end{equation}
	 Integrating over the height of the cloud column then gives the total kinetic energy in the cloud.

	 The power generated by one of these cloud columns (storms) depends upon the time required for a parcel of air to rise from the cloud base to the maximum cloud height ($\approx 80$ km). Utilizing the vertical velocities calculated in Figure~\ref{fig:CloudVel}, the cloud column rise time for Jupiter's moist convection is on the order of 45-135 minutes.
%	 For a given storm on Jupiter, observations place storm duration on the order of days \citep{Gierasch2000}. And for a given storm location, the time between storms is even longer. Therefore, the turnover time for the moist convection cycle detailed in Section 3 is on the order of weeks to months, or possibly longer. However, for a maximum power scenario, I assume the turnover time is 3 hours (cite source).
	 Given these times, Table~\ref{power}  lists the kinetic power outputs of each storm as a function of radius. The $\alpha$ parameter is held constant here at $0.2$. In addition, we also calculate the thermal power in the cloud in the same way as the kinetic, except our equation for thermal energy becomes (ignoring latent heat)
	 \begin{equation}
	 dQ = mC_p \Delta T = \rho\pi r^2 dz C_p (T_{cloud} - T_{environment})
	 \end{equation}
	 where $m$ is the mass, $C_p$ is the specific heat of water and $\Delta T$ is the difference in temperature between the cloud and the environment. These results are also listed in Table~\ref{power}. Regardless of cloud column radius, the kinetic power comprises $\sim 1.5\%$ of the total power.

	 \begin{table}[]
	 	\centering
	 	\caption{Kinetic and thermal power outputs for various cloud column radii using $\alpha=0.2$.}
	 	\label{power}
	 	\begin{tabular}{|c|c|c|c|}
	 		\hline
	 		Cloud Column Radius (km) & Kinetic Power (erg s$^{-1}$) & Thermal Power (erg s$^{-1}$)  \\ \hline
	 		1                        & $1.72 \times 10^{16}$                   & $1.28 \times 10^{18}$                                                     \\ \hline
	 		10                       & $2.26 \times 10^{19}$                   & $1.31 \times 10^{21}$                                                  \\ \hline
	 		25                       & $1.64 \times 10^{20}$                   & $9.27 \times 10^{21}$                                                     \\ \hline
	 		50                       & $6.88 \times 10^{20}$                   & $3.87 \times 10^{22}$                                                     \\ \hline
	 		100                      & $2.8 \times 10^{21}$                    & $1.57 \times 10^{23}$                                                   \\ \hline
	 		1000                     & $2.86 \times 10^{23}$                   & $1.6 \times 10^{25}$                                                     \\ \hline
	 	\end{tabular}
	 \end{table}

	 \citet{Gierasch2000} observed a Jovian moist convective storm 1000 km by 1000 km in area with a thermal power output of around $5\times 10^{22}$ erg s$^{-1}$ ($5\times 10^{15}$~W) as determined from infrared observations.  This value came from measuring the storm's thermal outflow within 1 scale height above 1 bar of pressure. Therefore, it is unclear what proportion of the total outflow energy comes directly from moist convection, and what comes from other sources like simple radiative cooling. The storm is likely comprised of many cloud columns of smaller radii, somewhere around 1-25~km as mentioned previously. Utilizing the thermal powers in Table~\ref{power} to match the observed power of the 1000 x 1000 km storm, 6 25-km radius cloud columns or $\sim 39,000$ 1-km radius cloud columns would be required. However, \citet{Gierasch2000} measured the thermal flux above 1 bar of pressure whereas our thermal powers are integrated over the entirety of the cloud column down to 6 bars. Correcting for this discrepancy and only utilizing the thermal power above 1 bar in our models, 20 25-km or $\sim 135,000$ 1-km cloud columns would be required.

	 \subsection{Oscillatory Mode Energies}\label{Mode_Energies}
	 Jovian thunderstorms have finite lifetimes and their energy must be dissipated into the near-surface environment. It is worth noting that currently no working theory exists as to exactly how the two the storm energy and mode excitation are coupled. The precise understanding of the nature of the coupling most likely requires the inclusion of atmospheric microphysics which is beyond the scope of this paper. However, on a more general scale, we hypothesize that the storm energy is transferred to the modes by the cloud column “punching” the upper atmosphere as it ascends. By analogy, this would be equivalent to a filled water balloon where someone flicks the balloon skin from the inside, thereby initiating the propagation of waves throughout the balloon. Assuming the energy can and does transfer to global oscillations, then in order to conclude whether moist convection could drive them, two key points must be understood: (1) how much energy is required to drive the modes (which, to first order, could be found by calculating the energy needed to produce an assumed surface displacement for a given oscillation mode and frequency); and (2) how the energy dissipation is partitioned into thermal transfer, radiation, winds, oscillations, etc. The latter is currently unknown, but the analysis of the former reveals it may not be necessary to understand in regards to oscillations.

	 Using a best-case scenario, we are interested in what the maximum possible energy supply can be from moist convection to Jupiter's global oscillations and whether that is sufficient to match the theoretical energies of the modes. Using the information in Table~\ref{power}, the maximum surface power density from moist convection kinetic energy is $\sim9 \times 10^{16}\,{\rm erg\,s^{-1}\,km^{-2}}$.  Utilizing 1997 Galileo data of Jovian thunderstorms as a tracer for moist convection, \citet{Gierasch2000} measure moist convective storm surface density on Jupiter to be $0.66 \times 10^{-9}$ km$^{-2}$. The surface area of Jupiter is $6.423 \times 10^{10}$ km$^2$ (using an equatorial radius at 1 bar of 71,492 km). Assuming this surface density is temporally typical and a storm surface area of $\sim 10^4$ km$^2$ is typical (\citet{Gierasch2000} specify these storms are typically a few hundred kilometers in diameter), then at any given moment there is approximately 42.4 moist convective storms occurring on Jupiter, or a total of $4.24 \times 10^5$ km$^2$ of storms. Using the maximum kinetic surface power density, moist convection could then supply a total of $\sim3.8 \times 10^{22}$ erg s$^{-1}$ to the global oscillations, assuming all available energy is used in the forcing. Keep in mind this value is a gross overestimate as not all of the observed storm area corresponds directly to a rising cloud column. As noted at the end of Section~\ref{maxenergypower}, a large storm will consist of several separated cloud columns and thus a large amount of the storm surface area does not contain a rising cloud column directly beneath it, such as $\sim 20$ 25-km radius cloud columns for a 1000 km$^2$ storm, which would only contribute $\sim 3.3 \times 10^{21}$ erg s$^{-1}$ to the modes.

	 \citet{Gudkova_Zharkov1999} calculate theoretical estimates of the energies for these Jovian global oscillations. For the fundamental $n=0$, $\ell=2$ mode, they find the energy in this mode per period is $6.4 \times 10^{26}\,{\rm erg}$ when a 1~m surface displacement is assumed. For the various models utilized to arrive at this value, the period ranges from 75 to 1600 minutes. The ideal scenario for driving oscillations would be a minimum mode energy, thus the longest period results in a power of $6.67 \times 10^{21}\,{\rm erg\,s^{-1}}$ for this particular mode.
	 
	 At first glance, it appears that there is enough kinetic power available to drive the modes. However, while the $n=0$, $\ell=2$ mode, for example, requires an average power of $6.67 \times 10^{21}\,{\rm erg\,s^{-1}}$ to survive, the storms do not produce the aforementioned kinetic powers on average, as the kinetic power is only produced over the rise time of the cloud column, or about 45 minutes for a 25-km radius cloud. If we assume a 1000 km x 1000 km storm erupts every day on Jupiter, then when we examine the total power required to drive just the $n=0$, $\ell=2$ mode, the kinetic energy from the storms falls two orders of magnitude short of the required driving energy.

	 In addition, the actual Jovian surface displacements for global oscillations are unknown. Thus, this is a free parameter in the calculation of mode energies. \citet{Gudkova_Zharkov1999} compute these energies assuming a surface displacement of 1~m. Using the shortest mode period, this translates to surface velocities of order $0.02\,{\rm cm\,s^{-1}}$. However, \citet{Vorontsov1976} and \citet{Bercovici_Schubert1987} predict global oscillations to have surface velocities of $10 - 100\,{\rm cm\,s^{-1}}$ For the 75 minute mode period, this results in  surface displacements on the order of $50 - 100$~m. Observational efforts place surface amplitudes at $49_{-10}^{+8}~{\rm cm\,s^{-1}}$ \citep{Gaulme2011}. Therefore, the 1~m estimate is a conservative estimate, adding to the disparity between moist convection kinetic energy and mode energies.

	 Given this analysis, if all the kinetic energy from moist convection does excite modes, the lowest fundamental mode possible to be excited would be of angular degree $\ell \approx$ few tens \citep[according to the predictions of][]{Vorontsov1976}, which would require all the energy to drive this singular mode, a mode with a surface amplitude much too small to observe. Higher-order radial modes could also be possible ($n>10$), again requiring the entirety of the moist convective kinetic energy. A far more likely scenario would be a partitioning of energy to excite several high order modes, yet likely none of these modes would be even close to observable limits. To reiterate, this scenario assumes maximum storm kinetic energy, minimum mode energy, 100\% energy conversion between the two, every square kilometer of a moist convective storm carries kinetic energy readily available to the oscillations, and a highly conservative surface displacement of 1 meter. Therefore, we can conclude that moist convective kinetic energy alone is insufficient to immediately excite any fundamental, low order, observable global oscillations.

	 \subsection{Kinetic Energy Partitioning}
	 As a final note, \citet{Bercovici_Schubert1987} estimate the power of turbulence from kinetic motions radiated into acoustic waves to be described by
	 \begin{equation}\label{BCEQ}
	 W = \frac{\rho u^3}{\lambda}\Big(M^3 + M^5 \Big)
	 \end{equation}
	 where $u$ is the eddy velocity, $\lambda$ is the eddy wavelength, and $M$ is the Mach number. If we use Equation~\ref{BCEQ} with our cloud models having velocities of 10-50 m s$^{-1}$ and the sound speed within Jupiter being $\sim 1$ km s$^{-1}$ we find that only $10^{-4}\% - 10^{-2}\%$ of the kinetic cloud energy actually gets converted to acoustic modes due to the very small Mach number ($10^{-6}$ to $10^{-4}$). If this is truly the case, then moist convective kinetic energy most certainly cannot excite Jovian modes.
	 
\section{Stochastic Harmonic Oscillator}\label{HarmonicOsc}
What we have shown with our model is that moist convective kinetic energy cannot immediately excite low-order oscillatory modes up to their expected energy and surface displacement. By analogy, we have shown that a parent of modest strength cannot push their child to maximum amplitude on a playground swing with only one push. Yet just as how the child can eventually reach a maximum swing amplitude from additional pushes at stochastic intervals, here we investigate if moist convective storms can stochastically interact with Jovian oscillations, eventually exciting them to observable amplitudes.

To investigate a stochastic situation, we create a 1D harmonic oscillation simulation for Jupiter.  The basic idea is simple: we begin with an interior model from which we compute a set of standard linear, adiabatic oscillation mode eigenfunctions and frequencies. We then set up several different schemes for how frequently and effectively moist convective storms deposit energy ``exciting'' the modes, given certain normalization conditions on the energy and surface displacement.  The modal energy is tracked as a function of time as it either dissipates or increases by encountering other storms with the appropriate frequency (parent pushing swing). After a sufficient amount of time, it is possible to determine if one or several modes attain enough energy to satisfy the theoretical estimates or observational lower limits. Below we discuss more details of these computations.

\subsection{Mode eigenfunctions}
% eigenfunction stuff
The modes we utilize in this simulation are a subset of eigenfunctions, both the radial and horizontal components, and their corresponding eigenfrequencies calculated from state-of-the-art Jupiter models \citep{Jackiewicz2012}. We only consider those with  angular degrees $\ell=0-15$  and radial orders $n=0-10$. The temporal eigenfrequencies span from  $\sim 0.1$~mHz to $\sim 2.5$~mHz, commensurate with the expectations of low-degree oscillations of Jupiter.
 
To establish a proxy for the initial energy of a given mode that takes into account its inertia we consider
\begin{equation}\label{E_and_Amps}
  E = \pi\omega^2 \int\limits_{0}^{R}\Big[\Big|\xi_r(r) \Big|^2 + l(l+1)\Big|\xi_h(r) \Big|^2 \Big]\rho r^2 \textnormal{d}r.
\end{equation}
Here, $\xi_r$ and $\xi_h$ are the radial and horizontal eigenfunctions, respectively, of the mode in question and $\rho$ is the radial density profile of the Jupiter model. As is common, the eigenfunctions are normalized at the surface according to the boundary conditions. At each timestep, once the energy imparted to each mode is computed, its radial (surface) amplitude  is determined according to 
\begin{equation}
  \xi_r(R) = A\tilde{\xi_r}(R) = A,
\end{equation}
where $\tilde{\xi_r}$ is the eigenfunction that is arbitrarily normalized to one at the surface. The ratio  between the radial and horizontal eigenfunction amplitudes is fixed by the solution of the oscillation equations. In what follows though, we only consider $A$, the surface displacement in real units of the mode in question. 

The eigenfunctions we use are not necessarily meant to be the most representative of the planet, as we've ignored higher-order effects due to rotation in the computation. As has been shown in similar efforts, rapid rotation causes subtle, but important changes to the eigenmode solutions of giant planets \citep{Fuller2014}. However, these functions provide a good approximation to how the oscillation inertia is distributed throughout the interior of the planet.

\subsection{Energy partition, growth, and decay}\label{energy}

Having computed the available kinetic energy from the storms  as discussed in Section~\ref{Mode_Energies}, we now specify when and how often the storms occur. Each storm is described by  two basic timescales, $t_{\rm load}$, and $t_{\rm storm}$. $t_{\rm storm}$ is the time over which the cloud column rises from the cloud base to its maximum height and $t_{\rm load}$ is the time between these rising cloud excitation events. A storm is triggered at $t=0$ and then each subsequent storm occurs at a later time randomly sampled between $t_{\rm load,max}$ and $t_{\rm load,min}$.

Each time a storm occurs, the energy available to the modes is the power from the cloud column(s) with a specified radius multiplied by $t_{storm}$. We assume energy equipartition where half of the available energy is transferred to thermal dissipation or bulk motions, or as \citet{Ingersoll2000} conclude, is returned to the South Equatorial Belt or Great Red Spot, and the other half is equally split between all the modes. However, the energy partitioned to a particular mode is not always added to the energy already contained within that mode. For any given mode excited with non-zero energy, we prescribe a phase, $\phi$, to it that oscillates sinusoidally in time between 0 and $2\pi$ at the mode's eigenfrequency.  So when a storm erupts, if $E_m$ is the energy partitioned to a particular mode, then the additional energy added to the mode is $E_m \cos\phi$. This  could result in modes being damped by some storms rather than amplified, thereby converting mode energy into thermal energy. This is a way to model the effect of ``pushing the swing'' at the wrong time. If at any point the mode's energy becomes negative it is reset to zero energy and the phase is reset to zero. 

We also run a suite of simulations for a distribution of storm energies. We create a Gaussian distribution with a mean of the minimum possible storm energy and a standard deviation equal to one-third of the maximum possible storm energy such that the maximum storm energy, or largest storm, will be three sigma away from the mean. When a storm erupts in the simulation, its energy is randomly chosen based on this  distribution and only considers values at or larger than the mean.

 To account for energy dissipation, we make use of the quality factor, $Q$, which parametrizes how well a particular mode ``rings'' within Jupiter. A  strongly frequency-dependent $Q$ is considered as well as a constant one. Following the work of  \citet{Mosser1995}, we use an approximate exponential relation given by 
\begin{equation}\label{Q}
  \log_{10}Q = 9  - 2\nu,
\end{equation}
in the range of $\nu=[0,3]$~mHz where $\nu$ is the temporal eigenfrequency of the mode.  For example, $Q\approx 10^7$ for a 1 mHz mode.  We also consider simulations where we simply fix  $Q$ at a constant value of $10^5$ for frequencies below 3~mHz  as suggested by \citet{Stevenson1983}. The Jovian atmospheric acoustic cutoff frequency has been estimated to be around 3.5~mHz, but the low-degree modes we consider are below this value.

Once values for $Q$ are determined for each mode, its energy loss over one oscillation cycle is given by
\begin{equation}
  Q = \frac{E}{\Delta E},
\end{equation}
where $E$ is its energy and $\Delta E$ is the energy lost.  $Q$ follows from either Equation~\ref{Q} or the constant relationship.
	 
\subsection{Simulation scenarios and setup}

This process allows us to create a time dependent map of the growth and/or decay of mode energies, mode amplitudes, and the overall surface displacement of the planet (i.e. the sum of the mode amplitudes). We then embed the code in a probabilistic distribution (see below) simulation which will run the entire process for a specified number of times and record the maximum energy achieved by each mode per run, the maximum amplitude of each mode per run, and the maximum planetary surface displacement per run. We run each simulation for 5000 days, more than adequate time for the oscillations to find a quasi-equilibrium based on various tests that were performed. Each simulation is run 200 times to build the probability distribution of maximum achieved mode energies and amplitudes. The final results allow us to thus analyze the mean of this stochastic process.
	 
Our overall strategy is put constraints on moist convection storms based largely upon theoretical estimates and observational limits of mode amplitudes and energies. In general, we consider the radial  $\ell=0$, $n=1$ mode as a proxy for all modes since it has the largest energy as its mode mass is the largest. If its theoretical energy prediction of $10^{26}$ - $10^{27}$~erg is failed to be met \citep{Vorontsov_Zharkov1981},  then we conclude  the specific parameters in that moist convection scenario to be unable to successfully drive modes. In conjunction, the cumulative surface displacement of the planet from all the modes shall be no smaller than 10 meters, a lower limit commensurate with the observed velocity amplitude of \citet{Gaulme2011}.

In the analysis we investigate four distinct moist convection scenarios, each of which reveals something about the possible nature of these storms and their relation to the modes.
\begin{enumerate}
\item It is likely that perhaps moist convective storms on Jupiter are much more frequent than we think. Small storms could be erupting from the cloud base all the time, yet they simply do not have enough buoyancy and upward momentum to reach observable cloud heights. We therefore investigate what temporal storm frequency is necessary for a 1-km radius storm with 100 columns per storm with a $t_{\rm storm}$ of 135 minutes (the time required for the cloud to reach maximum altitude from the cloud base) to achieve theoretical mode energy and amplitude estimates. This may also be thought of as 100 1-km radius storms erupting simultaneously across the planet.
	 	
\item What if the largest storms were primarily responsible for driving the modes? We thus also investigate the storm frequency necessary to achieve the theoretical mode energies and amplitudes for the largest physically feasible storms in our simulations, the 25-km radius cloud column with 200 columns per storm with a $t_{\rm storm}$ of 45 minutes.

\item We then examine the scenario where the energy in a storm contributing to the modes is actually much greater than just the kinetic energy that our cloud models suggest. Utilizing the storm energy found by \citet{Gierasch2000}, we find the storm frequency necessary to achieve theoretical mode estimates for storms of $5\times 10^{22}$ erg s$^{-1}$.

\item Finally we consider the probable case whereby a distribution of storm sizes can erupt. Using the one-sided Gaussian distribution of storms described in Section~\ref{energy}, we find what distribution of storm energies can reproduce theoretical mode energies and amplitudes.
\end{enumerate}
 	 
The values we use for $t_{\rm storm}$ are derived from the vertical velocities shown in Figure~\ref{fig:CloudVel}. For example, \citet{HuesoEtAl2002} show that a rising storm cloud column can have a radius no larger than 25 km, but that a moist convective storm can have up to 200 of these updrafts.   The 25 km radius column has an average updraft velocity of 30 m s$^{-1}$ over the 80 km range between the cloud base and maximum cloud height, resulting in $t_{\rm storm} = 45$~minutes. From Table~\ref{power}, we thus use an input power of $200\times 1.64\times 10^{20}\,{\rm erg\,s^{-1}} = 3.28\times 10^{22}\,{\rm erg\,s^{-1}}$. In addition, \citet{HuesoEtAl2002} allude to a storm frequency of about 12 days, thus we begin testing our simulations with the minimum and maximum values of $t_{\rm load}$ to be 9 and 15 days, respectively. The value for $t_{\rm load}$ is randomly sampled between its maximum and minimum values, that is, once a storm concludes the time until the subsequent storm erupts is randomly selected to be between 9 and 15 days, for example. This temporal range is then adjusted between simulations to find which storm eruption frequencies produce the mode energies that agree with theoretical estimates and observed amplitudes.

	 \section{Harmonic Oscillation Results}\label{HarmonicResults}

	 \subsection{Case 1}\label{Case1}
	 In the scenario in which only the smallest storms contribute to driving Jovian modes, we find that for a 1 km radius storm with 100 columns per storm with a $t_{\rm storm}$ of 135 minutes, there is no temporal spacing between storm eruptions that reproduces the desired mode energies given the predicted kinetic energy available in these columns. If one of these storms erupts every 1-2 days, the $l=0$, $n=1$ mode has an energy of $2\times 10^{21}$ ($1.5\times 10^{21}$) erg and Jupiter has an overall surface displacement of 10.5 (5.3) cm for the frequency-dependent (constant) quality factor. In order to achieve theoretical energies, it would require an eruption of $\sim 10^7$ single column storms or $\sim 10^5$ 100 column storms each day on Jupiter. This would imply an average surface density of one storm column per 6423 km$^2$. 
	 
	 \subsection{Case 2}
	 When examining the case of the largest realistic storm frequency, we reach a similar conclusion to Case 1. For a storm of 200 columns with each column having a radius of 25 km, having one of these storms erupt every 1-2 days, the $l=0$, $n=1$ mode has an energy of $1.4\times 10^{25}$ ($9\times 10^{24}$) erg and Jupiter has an overall surface displacement of 7.3 (4.1) meters for the frequency-dependent (constant) quality factor. Theoretical estimates would require $\sim 70$ of these storms to erupt per day. However, given that \citet{HuesoEtAl2002} allude to a $\sim 12$ day spacing of these storms, coupled with the observations made by \citet{Gierasch2000} which indicate an average abundance of $\sim 43$ of these storms being present on Jupiter at any given time (not all of which are as large as the one observed; see Section~\ref{Mode_Energies}), we conclude that there are simply too few storms of this size to drive oscillatory modes.
	 
	 \subsection{Case 3}
	 Perhaps our derivations of the available storm energy are incorrect. Perhaps there is actually much more storm energy readily available to the modes than just the kinetic energy we have considered. Or maybe there is additional energy associated with Jovian storms that can also contribute to mode driving. As an upper limit we consider the entirety of the storm energy observed by \citet{Gierasch2000}. They observed a storm with a power of $5\times 10^{22}$ erg s$^{-1}$. Using their estimated storm surface density of $0.66 \times 10^{-9}$ km$^{-2}$, we find that storms like this would supply $\sim 3.25 \times 10^{30}$ erg day$^{-1}$ to the modes (found by $\frac{5\times 10^{22}\ erg\ s^{-1}}{10^6\ km^2} \div 0.66 \times 10^{-9}\ km^{-2}$ converted to erg day$^{-1}$ and halved as we assume only half the available storm energy drives modes). When utilizing this energy input in our simulations, we find the mode energies and amplitudes are much too high, resulting in an energy of $3.5 \times 10^{29}$ erg for the $l=0$, $n=1$ mode and an overall surface amplitude of 775 meters.
	 
	 Although this does not reproduce the mode estimates we expect, it is promising because it does reveal that enough associated storm energy is indeed present to excite Jovian modes via moist convection. By varying the daily energy input from the storms, we find that in order to excite the modes to their theoretical energies and observed amplitudes ($10^{26} - 10^{27}$ erg for the $l=0$, $n=1$ mode and a 10-100 meter cumulative surface amplitude), we find that the modes require, on average, a daily storm eruption that provides between $5 \times 10^{27}$ and $10^{28}$ erg day$^{-1}$. If all the observed energy in the storm were available to modes, then this energy requirement could be achieved by an average surface abundance of approximately 43 Jovian storms of average surface area 764 km$^2$, or approximately 33 storms with average surface area 1000 km$^2$. It is worth noting that even though these storm abundances are commensurate with those in Section~\ref{Mode_Energies}, these storms utilize the total thermal surface power density as that observed by \citet{Gierasch2000}, not the kinetic power.
	 
	 \subsection{Case 4}
	 Using a Gaussian distribution of storm energies  as described in Section~\ref{energy}, we investigate what distribution of storm energies can excite modes to predicted levels of energy and amplitude. If the minimum energy of the distribution (recall the minimum energy is the mean because it is a one-sided Gaussian) matches our smallest considered storm energy of $8.1\times 10^{20}$ erg with a standard deviation one-third the maximum considered storm energy of $1.35 \times 10^{26}$ erg (so that the largest storms are three standard deviations from the mean), we find the $l=0$, $n=1$ mode energy to have $1.5 \times 10^{25}$ ($6\times 10^{24}$) erg for the frequency dependent (independent) quality factor and an overall surface amplitude of 7.6 (3.8) meters. So for a distribution of storm energies contained within the kinetic energy limits of our cloud models, we find that there is not enough energy to supply to the modes.
	 
	 However, we do find two distributions that can successfully reproduce the mode predictions. If both the mean and standard deviation are 100 times larger, such that the minimum energy is $8.1\times 10^{22}$ erg and three sigma from the mean is $1.35 \times 10^{28}$ erg, then the overall surface amplitude of Jupiter becomes 38 meters and the $l=0$, $n=1$ mode has an energy of $8\times 10^{26}$ erg. The storm cadence for this simulation was an eruption every 1-2 days, although increasing this temporal range only slightly decreases the resultant amplitudes and energies due to the large quality factor. 
	 
\begin{figure}[t]
	 	 	\centering
	 	 	\centerline{
	 	 		\includegraphics[width=.45\textwidth]{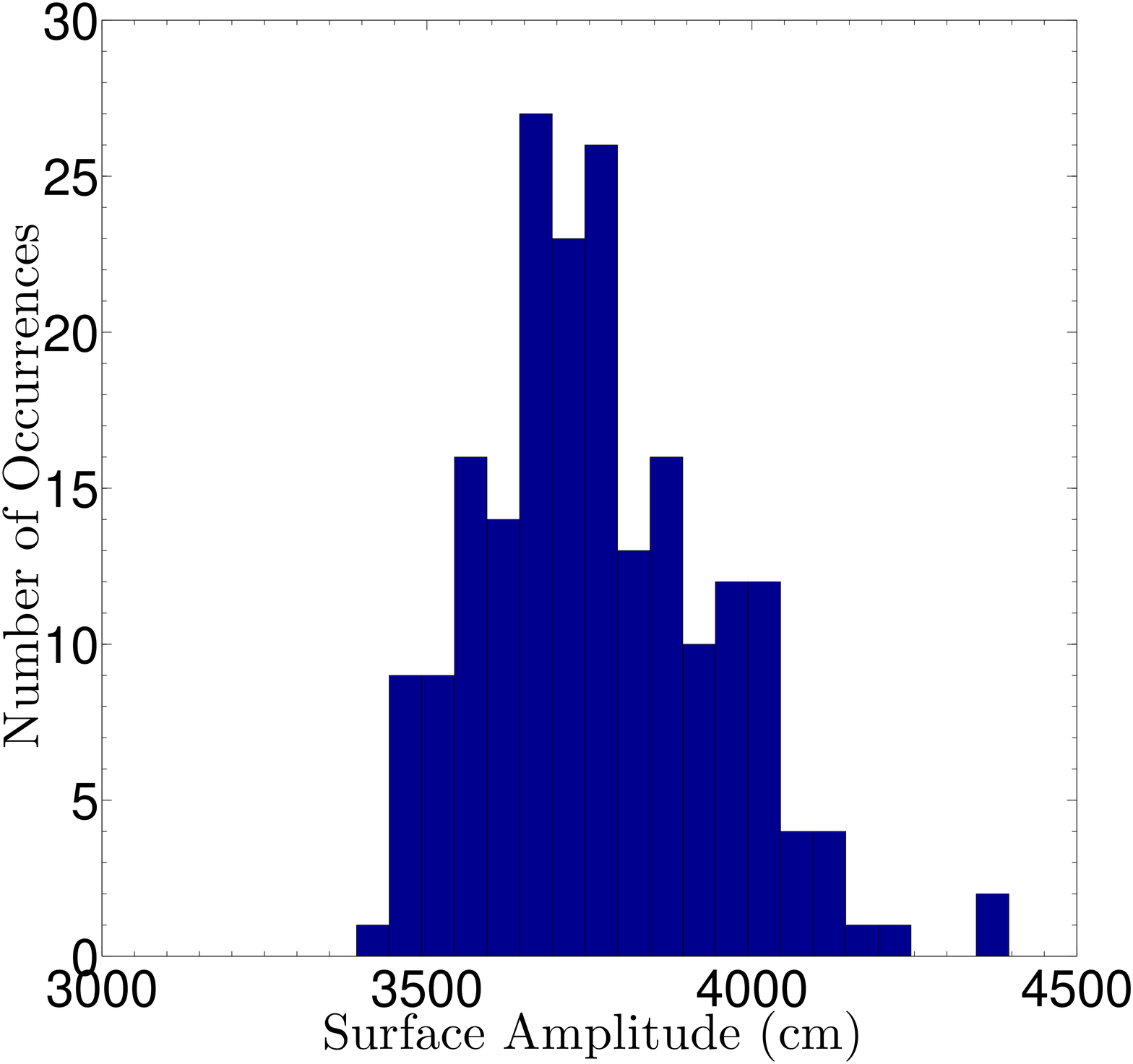}
	 	 		\includegraphics[width=.5\textwidth]{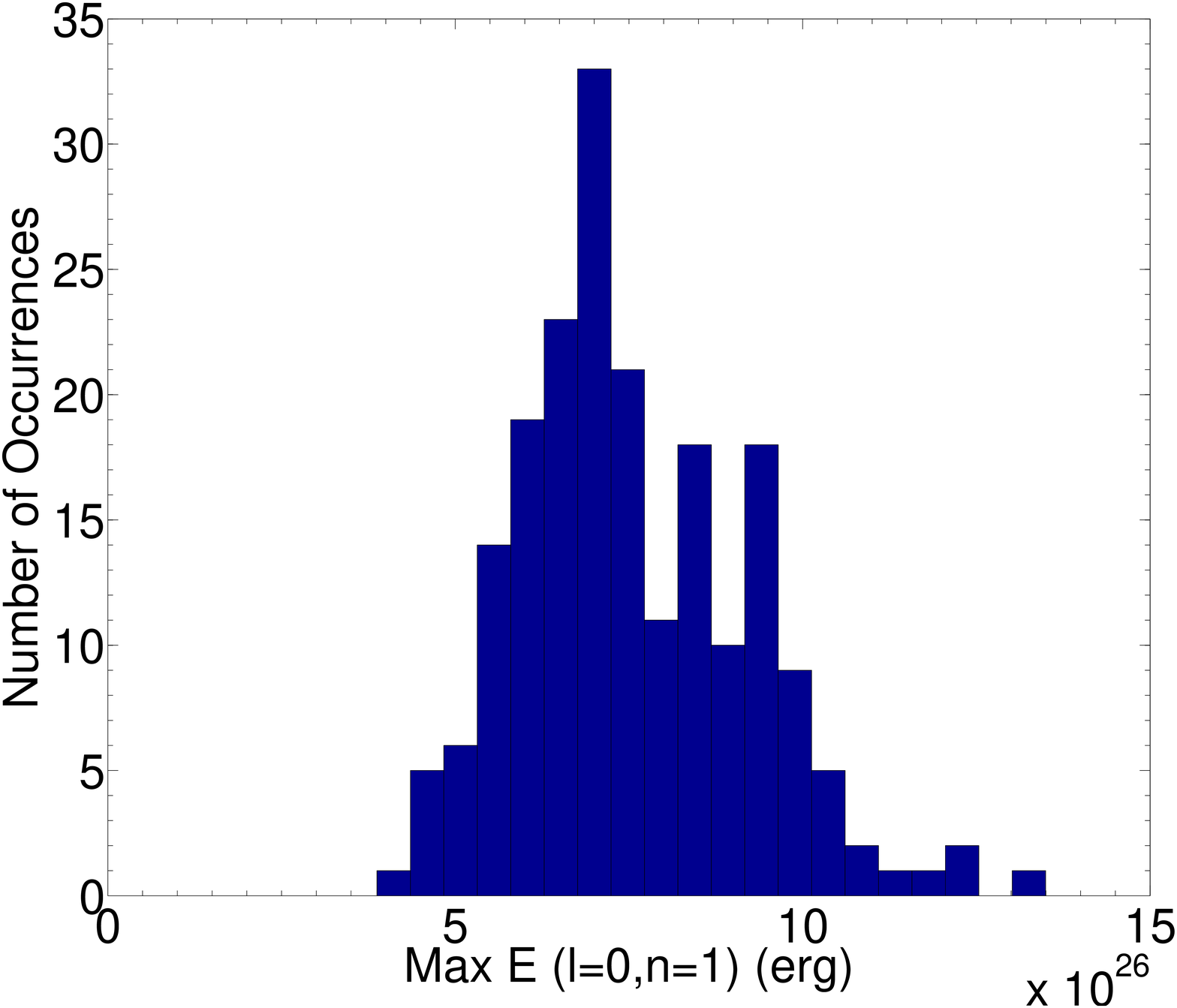}}
	 	 	\caption{The overall surface displacements and $l=0$, $n=1$ mode energies for the Case 4 simulation with minimum storm energy of $8.1\times 10^{20}$ erg and a standard deviation of $1.35 \times 10^{28}$ erg. Left: The maximum cumulative surface displacement of all modes achieved in each simulation run and how many times that displacement was reached. Right: The maximum energy in the $l=0$, $n=1$ mode reached in each simulation run and how many times that energy was reached. The simulation was run 200 times to produce these distributions.}
	 	 	\label{fig:Case4}
	 	 \end{figure}

	 The results of the second simulation are shown in Figure~\ref{fig:Case4}. For this energy distribution the minimum energy is still that of our smallest storm ($8.1\times 10^{20}$ erg), but with a standard deviation 100 times larger than our initial estimate ($1.35 \times 10^{28}$ erg). For a storm cadence of one eruption every 1-2 (5-10) days the $l=0$, $n=1$ mode contains an average energy of $8.2\times 10^{26}$ ($3.2 \times 10^{26}$) erg with an overall surface displacement of 38 (29) meters. The spread in energies and amplitudes in Figure~\ref{fig:Case4} is typical of these simulations.
	 
	 \section{Discussion}\label{Discussions}

\begin{figure}[t]
	 	\centering
	 	\includegraphics[width=.5\textwidth]{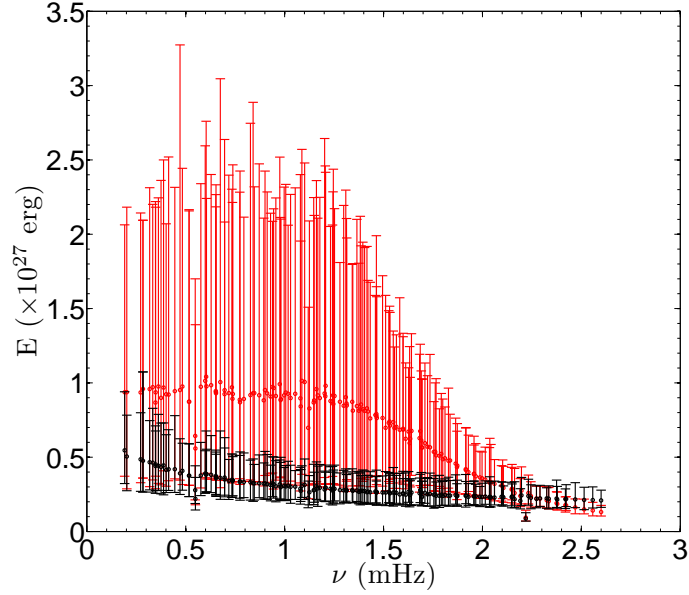}
	 	\caption{Distribution of the maximum energies obtained by the full set of eigenmodes in 200 simulations for Case 4. The red lines correspond to the simulations run with a frequency-dependent $Q$ and the black lines for the constant $Q$ case. In both cases, for each of the 200 simulations we determine the maximum energy achieved by each individual mode throughout the simulation. Thus, for each vertical line in the figure, the filled circles represent the average of the 200 energy maxima for each mode, the top bar the absolute maximum from all 200 simulations, and the bottom bar the absolute minimum.}
	  	\label{fig:Qcomp}
	 \end{figure}

\begin{figure}[t]
 		\centering
 		\centerline{
 		\includegraphics[width=.5\textwidth]{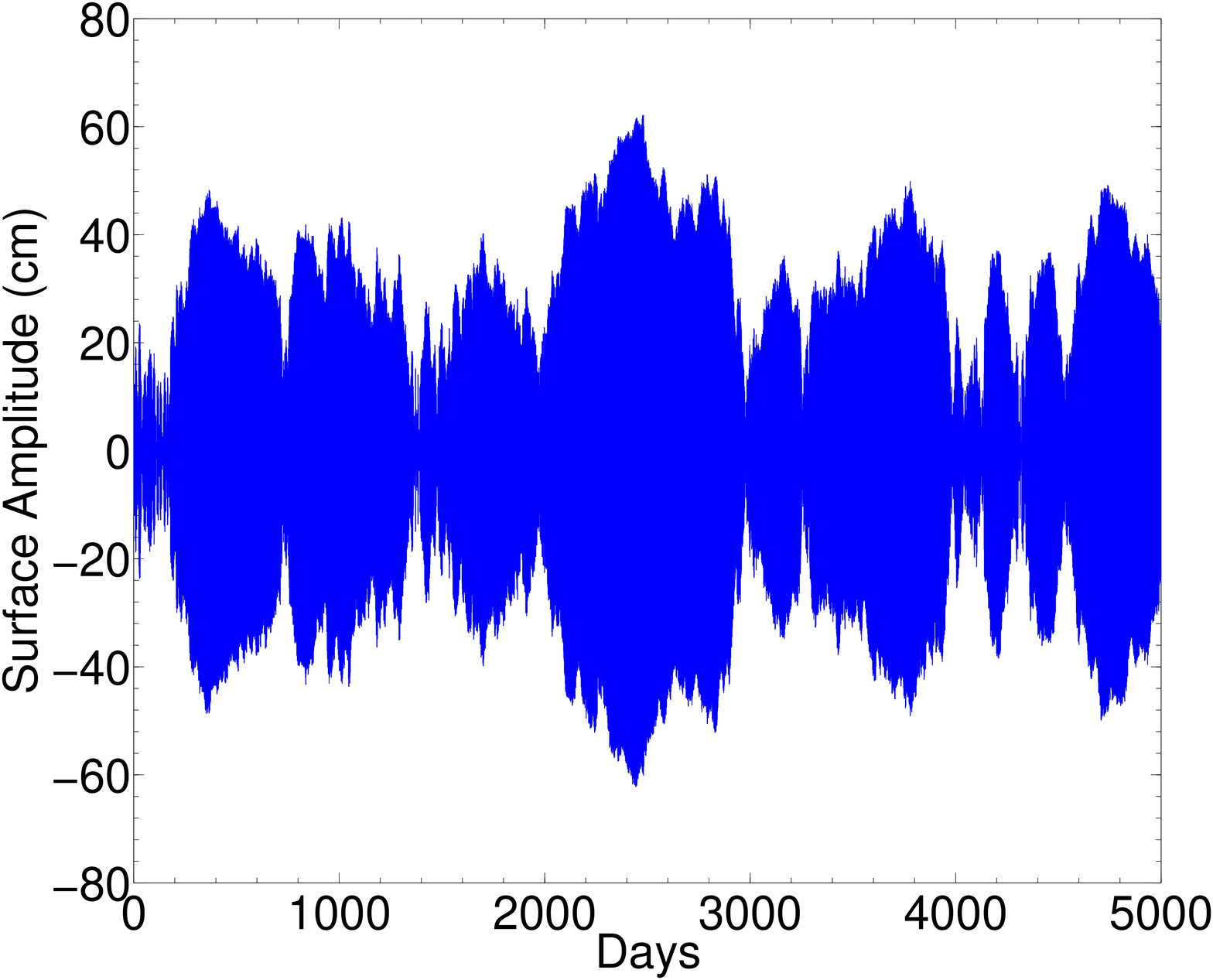}
 		\includegraphics[width=.5\textwidth]{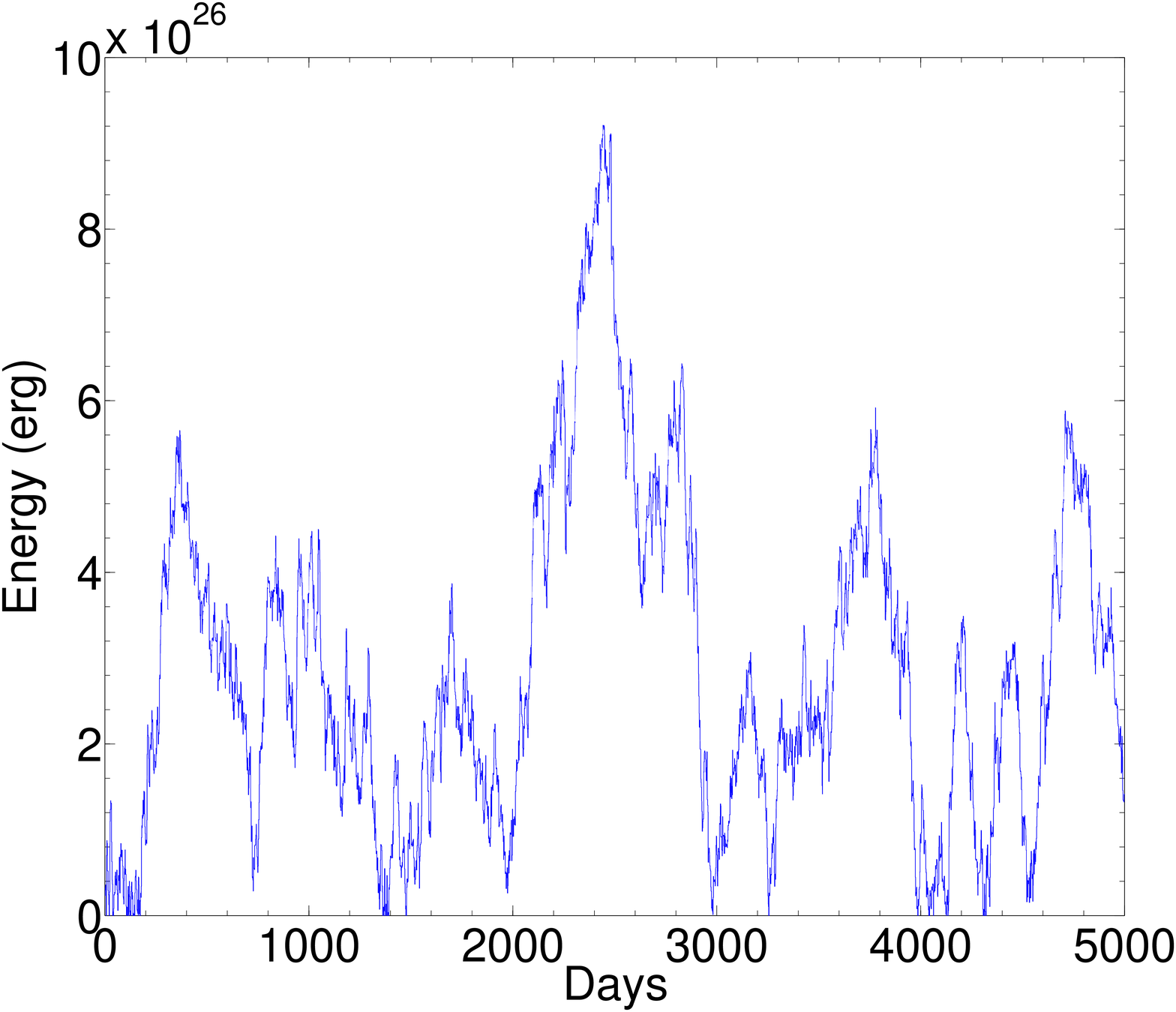}}
 		\caption{Left: An example of the surface amplitude as a function of time for  the $l=0$, $n=1$ mode from one of the 200 runs of the Case 4 simulation. Right: An example of the energy lifetime of the same mode in the same simulation.}
 		\label{fig:l0n1}
 		\end{figure}

	 Regardless of which simulation case was investigated, we find that the results of the $l=0$, $n=1$ mode energy differ by a small amount regardless of which quality factor is used, even though the higher frequency modes may differ more so between quality factors. However, we do find the frequency-dependent $Q$ produces results that are closer to observations.  For the exact relationship between $Q$ and frequency in Equation~\ref{Q}, any modes with frequency greater than 2.5 mHz are very quickly damped. Therefore, the frequency dependent $Q$ is a more realistic representation for Jupiter as the higher frequency modes decrease in amplitude more quickly than in the frequency independent Q case, consistent with Figure 2 in \citet{Gaulme2011}. This can be seen in Figure~\ref{fig:Qcomp}. (Note: in Figure~\ref{fig:Qcomp} we expect the $l=0$ $n=1$ mode, which has the lowest frequency, to have the greatest maximum energy, but the modes that exceed the energy of the $l=0$ $n=1$ mode are simulation outliers with the majority of their maximum mode energies below that of the $l=0$ $n=1$ mode.) Radial orders of $n=4$ and greater, corresponding to frequencies around 2.75 mHz, were observed by \citet{Gaulme2011}, but unfortunately our subset of modes utilized in this work do not extend up to that frequency. In addition, neither quality factor can account for the missing mode amplitudes near 2 mHz, as both predict the survival of modes near this frequency. Therefore, either the observations made by \citet{Gaulme2011} were made during a period over which the $\sim 2$ mHz modes were being temporarily damped, energy equipartition does not favor modes of this frequency, or the Jovian quality factor is more complex than we think. A clearer picture of mode survival is illustrated in Figure~\ref{fig:l0n1}. It is clear that due to the stochastic nature of mode excitation and de-excitation from moist convection, there are periods in which mode amplitudes are suppressed, or even drop to zero, thereby lending credibility to the idea that \citet{Gaulme2011} observed Jupiter during a period over which the $\sim 2$ mHz modes were being temporarily damped.
	 
	 Table~\ref{SelectedResults} lists the simulations with the most promising results for achieving theoretical mode energies and amplitudes for each of the four cases. The simulations that produce results which agree with theoretical estimates are highlighted in green. Assuming that moist convection is solely responsible for driving Jovian modes, we find the kinetic energies available as calculated from our cloud models  insufficient to do so. There are several possible reasons for this:
	 \begin{itemize}
	 	\item The quality factor on Jupiter is actually much higher than previously thought. A much greater $Q$ would prevent modes from decaying as quickly. However, a quality factor greater than $10^5$ is unlikely to have much of an effect on mode survival as can be seen in Figure~\ref{fig:l0n1}. It is actually the destructive interference of storm excitation and the current mode phase that is responsible for temporarily driving the mode energies down to zero. $Q$ is large enough that little energy is lost between storms, and feasible as shown by \citet{Stevenson1983}.
	 	\item Not all the excitation energy for Jovian oscillations comes from moist convection. That is, a secondary or even tertiary energy source could also contribute to the modes. Turbulent convection, albeit at a much lower energy scale than the sun, does still exist within Jupiter, possibly also contributing to mode excitation. If a secondary excitation source does exist, however, it must be located near the surface of Jupiter as the inertia becomes increasing difficult to overcome at decreasing radii.
	 	\item We may lack proper understanding of energy equipartition. Perhaps more or less energy is allocated to the modes or there is some mechanism that preferentially favors the excitement of some modes over others, such as in the observed case of 2 mHz suppression by \citet{Gaulme2011}.
	 	\item The energy from the storms is underestimated. It is certainly possible that our cloud models lack a key source of energy associated with them that could also be transferred to the modes, including thermal energy.
	 	\item Our harmonic oscillation code may be overly simplistic and therefore is a poor representation of the complex interaction between moist convection and the oscillations. To better understand exactly how energy is transferred to the modes, one would need a simulation that accounts for Jovian atmospheric microphysics.
	 \end{itemize}
 
 	Finally, if we allow for mode excitation to be a byproduct of more storm energy than just kinetic, our simulations predict that it is in fact the largest, most energetic storms responsible for driving Jovian modes. This can be inferred in two ways. First, it can be seen in Figure~\ref{fig:l0n1} that mode growth is not indefinite. Due to the stochastic nature of the excitation and de-excitation, the energy in a particular mode is likely to return to zero at times. As such, little energy input from small storms cannot slowly drive up mode energies to theoretical estimates, even for a large quality factor. The second way this can be inferred is from the simulations in Case 4. Using a distribution of storm sizes to excite the modes, theoretical estimates can only be achieved if the standard deviation of the distribution extends into high enough storm energies. This tells us that if we want to achieve highly energetic modes, we need highly energetic storms. This could be excellent news when trying to explain mode excitation on Saturn, as it has extremely large quasi-periodic moist convective storms \citep{Li_Ingersoll_2015}.

 	\begin{table}[]
 		\centering
 		\caption{Summary of selected results. The Case corresponds to the simulation cases in Section~\ref{HarmonicResults}, Q is the quality factor (D=frequency Dependent, I=frequency Independent), t$_{storm}$ is the duration of the storm, P$_{storm}$ is the power in the storm, SE$_{min}$ and SE$_{max}$ are the minimum storm energy and 3 sigma maximum storm energy in the storm Gaussian energy distribution, respectively, SA is the maximum cumulative surface amplitude of Jupiter from all modes, and E is the maximum energy in the $l=0$ $n=1$ mode. For Cases 1 and 2, t$_{load}$ is 1-2 days.}
 		\label{SelectedResults}
 		\begin{tabular}{|c|c|c|c|c|c|c|c|}
 			\hline
 			Case & Q & t$_{storm}$ (min) & P$_{storm}$ (erg/s) & SE$_{min}$ (erg) & SE$_{max}$ (erg) & SA (m) & E (10$^{26}$ erg) \\ \hline
 			1    & D & 135            & $10^{18}$             & N/A               & N/A               & 0.105           & 2$\times10^{-5}$                          \\ \hline
 			1    & I & 135            & $10^{18}$             & N/A               & N/A               & 0.053          & 1.5$\times10^{-5}$                          \\ \hline
 			2    & D & 45             & 3.24$\times 10^{22}$             & N/A               & N/A               & 7.3             & 0.14                         \\ \hline
 			2    & I & 45             & 3.24$\times 10^{22}$             & N/A               & N/A               & 4.1            & 0.09                          \\ \hline
 			3    & I & 45             & 3.76$\times10^{25}$          & N/A               & N/A               & 775             & 3500                          \\ \hline
 			3    & D & 45             & 3.76$\times10^{25}$          & N/A               & N/A               & 1600            & 6000                          \\ \hline
 			\rowcolor[HTML]{34FF34} 
 			3    & I & 45             & 5.79$\times10^{22}$          & N/A               & N/A               & 31              & 5.5                           \\ \hline
 			\rowcolor[HTML]{34FF34} 
 			3    & D & 45             & 5.79$\times10^{22}$          & N/A               & N/A               & 63              & 10                            \\ \hline
 			4    & D & 45             & N/A              & 8.1$\times10^{20}$            & 1.35$\times10^{26}$           & 7.6             & 0.15                          \\ \hline
 			4    & I & 45             & N/A              & 8.1$\times10^{20}$            & 1.35$\times10^{26}$           & 3.8             & 0.06                          \\ \hline
 			\rowcolor[HTML]{34FF34} 
 			4    & I & 45             & N/A              & 8.1$\times10^{22}$            & 1.35$\times10^{28}$           & 38              & 8                             \\ \hline
 			\rowcolor[HTML]{34FF34} 
 			4    & I & 45             & N/A              & 8.1$\times10^{22}$            & 1.35$\times10^{28}$           & 41              & 8.1                           \\ \hline
 		\end{tabular}
 	\end{table}

	\subsection{Water Abundance on Jupiter}\label{waterabundance}
	As a final note, it is worth examining the assumption made in Section~\ref{CloudModel} regarding the water abundance on Jupiter. Although we do not quantitatively examine the effect of a varying atmospheric water abundance, an abundance variation does indeed affect the moist convective process and therefore our results. If the water abundance in Jupiter's atmosphere is decreased, the water cloud base would form at a lower pressure. For example, if the water abundance were 1 percent that of solar as suggested by \citet{Bjoraker1985}, the cloud base would form around 2.5 bars of pressure \citep{Stoker1986}. This would result in lower cloud column momentum, and thus less kinetic energy available to drive oscillations.
	
	Conversely, if the water abundance is greater than three times solar we would expect to see more energy available to the modes for reasons opposite that described above, until an enrichment of 9.9 times solar is reached. \citet{Leconte_etal2017} and \citet{Guillot1995} show that for a water enrichment greater than 9.9 times solar, the atmosphere becomes more stabilized against convection, thereby inhibiting storms. The result is a more time-dependent moist convection. Although this may explain the quasi-periodic nature of Saturn's moist convective storms as described in \citet{Li_Ingersoll_2015}, Jupiter's moist convection is more stochastic, thereby limiting the water abundance estimate to no more than 10 times solar. Since we assumed an enrichment of 3 times solar for this work and the range of Jupiter's water abundance is approximately limited between 1 and 10 times solar, it is safe to say that the kinetic energy available to modes from moist convection is on the conservative estimate of the water abundance spectrum, thereby making it more likely that moist convection can indeed drive Jupiter's oscillations. We did not run simulations with a varying water abundance as the kinetic energy is so low relative to the storm energy observed by \citet{Gierasch2000} that even an increase of an order of magnitude of available kinetic energy would still fall short of the required mode excitation energy.

	 \section{Conclusions}\label{Conclusions}
	 Here we have shown that moist convection does not have enough kinetic energy in the rising cloud columns to immediately excite Jovian oscillations to predicted levels of energy and surface amplitude. Therefore, if moist convection is truly the excitation mechanism responsible for driving Jupiter's modes, the modes must, at minimum, be excited stochastically by moist convective storms such that the storms can continually provide energy to the modes. However, even with stochastic excitation, we find that our storm cloud column models cannot provide enough kinetic energy to achieve theoretical mode energies and amplitudes regardless of quality factor and regardless of the time between storm excitations. The modes can be sustained at theoretical energies, however, given a supply of $5\times 10^{27}$ to $10^{28}$ erg per day. We find that from the observations of \citet{Gierasch2000}, there is more energy in Jupiter's large storms than the theoretical mode energy estimates. Therefore, should moist convection truly be responsible for mode excitation, then this is indicative of more storm energy being available to the modes than previously thought.
	 
	 We also find that the frequency dependent and independent quality factors can both reproduce similar energies in the low frequency modes (around 1 mHz) but both fail to reproduce the damped modes around 2 mHz as observed by \citet{Gaulme2011}. Yet, those observations could have been recorded at a time in which the 2 mHz modes happened to be temporarily damped by the moist convection, or energy equipartition simply does not favor those frequencies.
	 
	 Finally, we find it is the largest storms with the largest energies that are primarily responsible for mode excitation. If the required mode energy did come from the kinetic energy in rising cloud columns, and if storms with a cloud column radius of 1 km are instead responsible, it would require the average existence of one column per 6423 km$^2$ on Jupiter at any given moment. Currently no observations exist to support this hypothesis, however given that columns of this size would only rise to 3 bars of atmospheric pressure or less, they may not be able to be directly observed at all. Similarly, for the case of storms comprised of 200 cloud columns, each with a 25 km radius, theoretical mode estimates would require $\sim 43$ of these storms to erupt per day, something inconsistent and far more frequent than \citet{HuesoEtAl2002} and \citet{Gierasch2000} would suggest. Overall, it appears that moist convection is not responsible for driving Jovian oscillations, unless more storm energy other than just the kinetic in the rising cloud column is available to the modes.

         \bibliography{All_Citations_MC}

\end{document}